\documentclass[final,5p,times,twocolumn,amssymb]{elsarticle}
\usepackage{graphicx}
\usepackage{dcolumn}
\usepackage{bm}
\journal{PHYSICA E} \tabcolsep1.49mm
\begin{document}
\title{In-Plane Magnetoresistance on the Surface of Topological Insulator}
\author{Morteza Salehi}
\author{Mohammad Alidoust}
\author{Yousef Rahnavard}
\author{Gholamreza Rashedi}
\address{Department of Physics, Faculty of Sciences, University of Isfahan,
81744 Isfahan, Iran}
\begin{frontmatter}
\begin{abstract}
We study the tunneling magneto-transport properties of the
Ferromagnetic Insulator-Normal Insulator-Ferromagnetic
Insulator(F$\mid$N$\mid$F) and Ferromagnetic Insulator-Barrier
Insulator-Ferromagnetic Insulator (F$\mid$B$\mid$F) junctions on
the surface of topological insulator in which in-plane
magnetization directions of both ferromagnetic sides can be
parallel and antiparallel. We derive analytical expressions for
electronic conductances of the two mentioned junctions with both
parallel and antiparallel directions of magnetization and using
them calculate magnetoresistance of the two junctions. We use thin
barrier approximation for investigating the F$\mid$B$\mid$F
junction. We find that although magnetoresistance of the
F$\mid$N$\mid$F and F$\mid$B$\mid$F junctions are tunable by
changing the strength of magnetization texture, they show
different behaviors with variation of magnetization. In contrast
to the magnetoresistance of F$\mid$N$\mid$F, magnetoresistance of
F$\mid$B$\mid$F junctions shows very smooth enhance by increasing
the strength of magnetization. We suggest an experimental set up
to detect our predicted effects.
\end{abstract}

\end{frontmatter}

\section{Introduction}
\label{intro} Recently, some topological insulators (TI) are
observed experimentally and investigated theoretically
\cite{Hsieh,Bernevig,Koenig,Kane,Moore,Xia,Hor,Fu1,Fu2,linder1,linder2,Zhang,
Raghu}.
 A TI is an insulator in bulk but a robust conducting in edge even in
presence of deformation or strong disorder \textit{i.e.} it is
gapless only in surface. So it can carry charge current on its
surface even in the absence of magnetic field. This state of
condensed matter is due to spin-orbit coupling and protected by
origin of time-reversal symmetry \cite{Fu1,Fu2,Moore}.  For
two-dimensional TI we have helical modes (the spin orientation of
surface electrons are perpendicular to momentum direction ) at the
edges \cite{Bernevig}, similar to quantum Hall systems.
\begin{figure}[tbp]
\centering{\includegraphics[width=0.8\columnwidth]{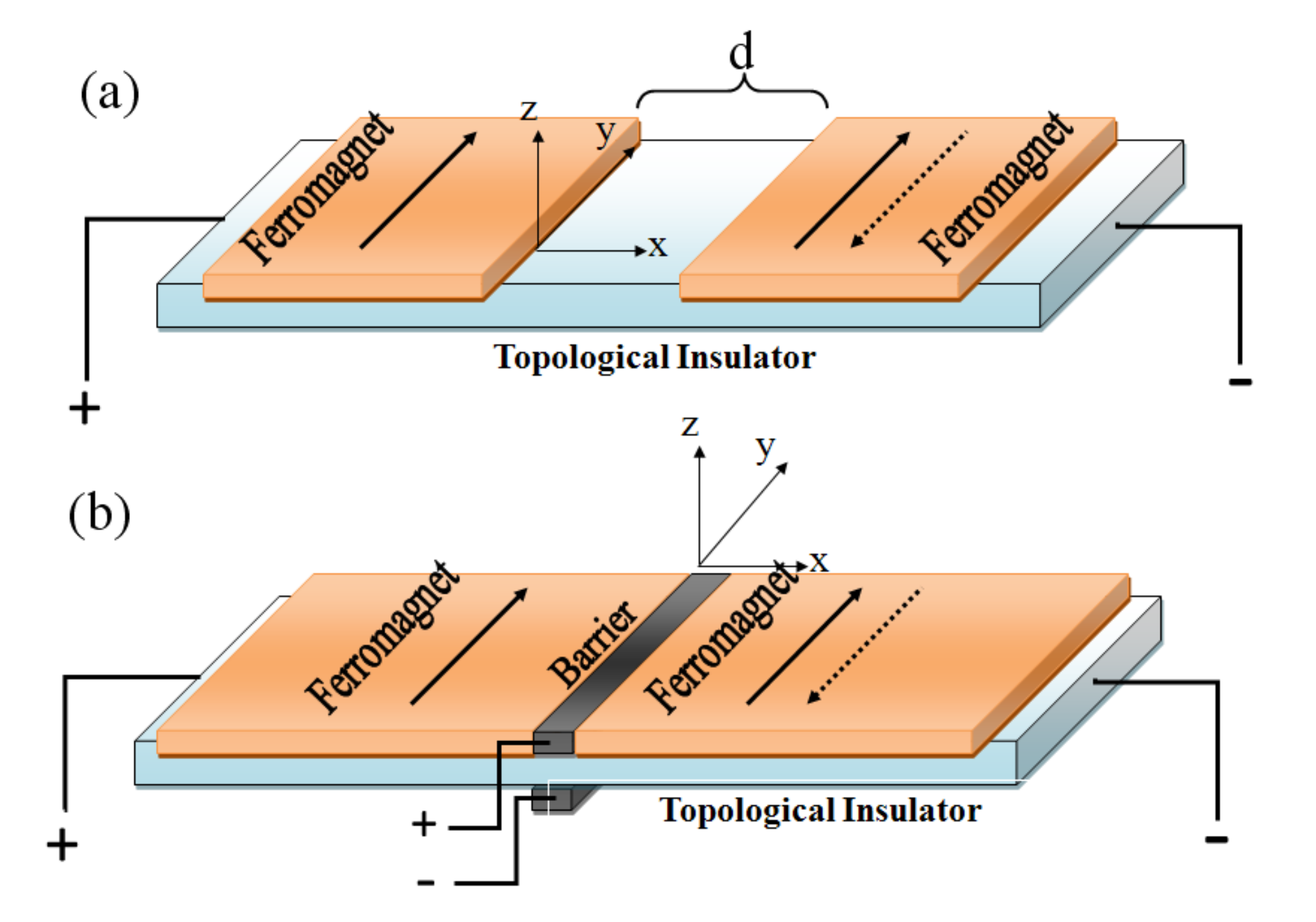}}
\caption{(Color online) Schematic model of the suggested set up for
$a$) F$\mid$N$\mid$F and $b$) F$\mid$B$\mid$F junctions on top of
the topological insulator. We assume that middle of the junctions is
located at $x=d/2$ and directions of the magnetizations in both
sides can be parallel and antiparallel. For F$\mid$B$\mid$F case we
employe thin barrier approximation namely, when transverse voltage
$V_0 \rightarrow\infty$ and $d\rightarrow\infty$.} \label{Model}
\end{figure}
\begin{figure}[tbp]
\centering{\includegraphics[width=0.8\columnwidth]{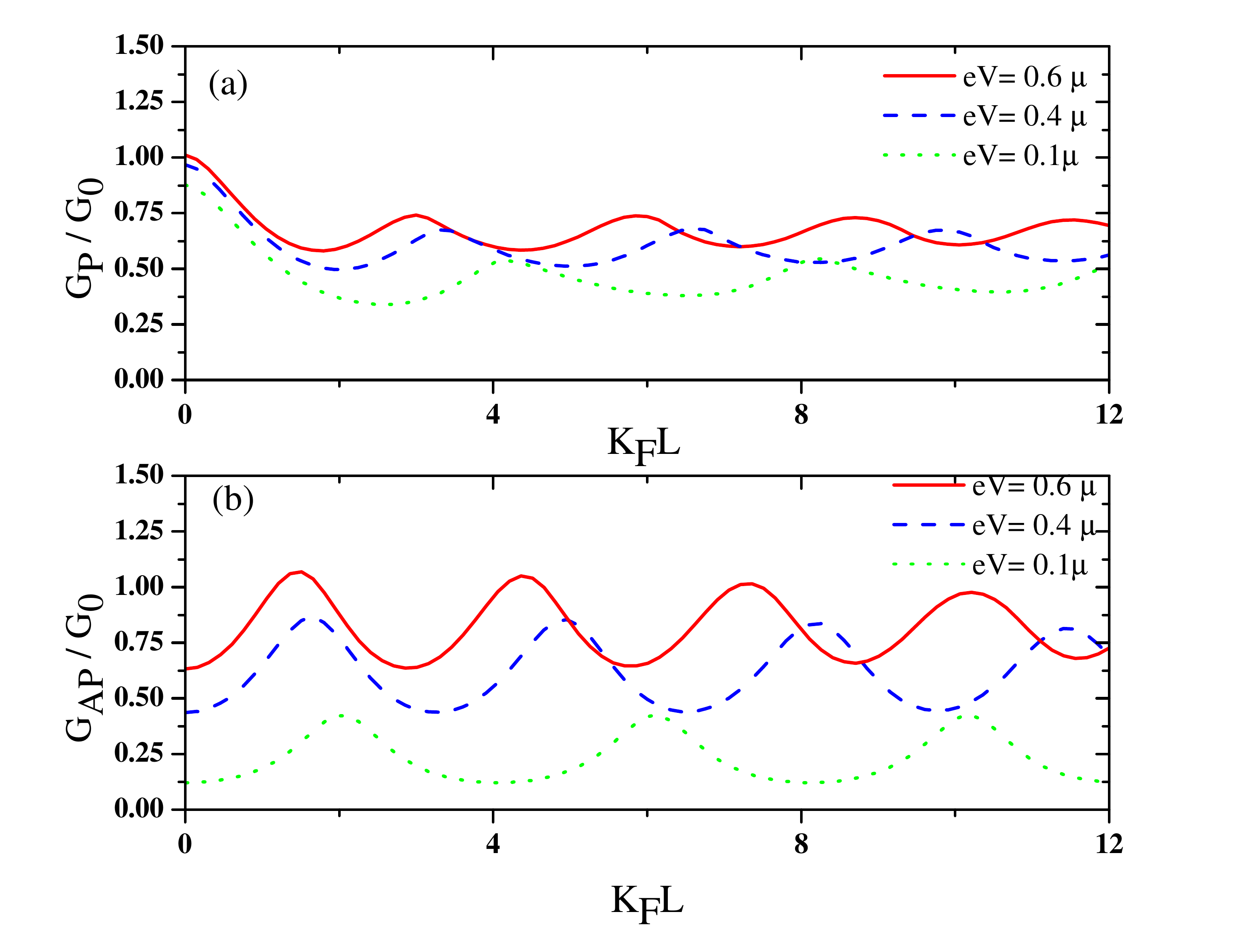}}
\caption{(Color online) The electronic conductances for parallel
($a$) and antiparallel ($b$)
 directions of magnetization in F$\mid$N$\mid$F junctions. Here we have
 fixed the $M/M_C=0.9$ and plotted electronic conductances vs. width of the
 normal region ($K_FL=\frac{\mu}{\hbar v_F}L$) for three values of $eV$ gate voltage.  } \label{[I]GP-GAP-kfL}
\end{figure}
\begin{figure}[b]
\centering{\includegraphics[width=0.8\columnwidth]{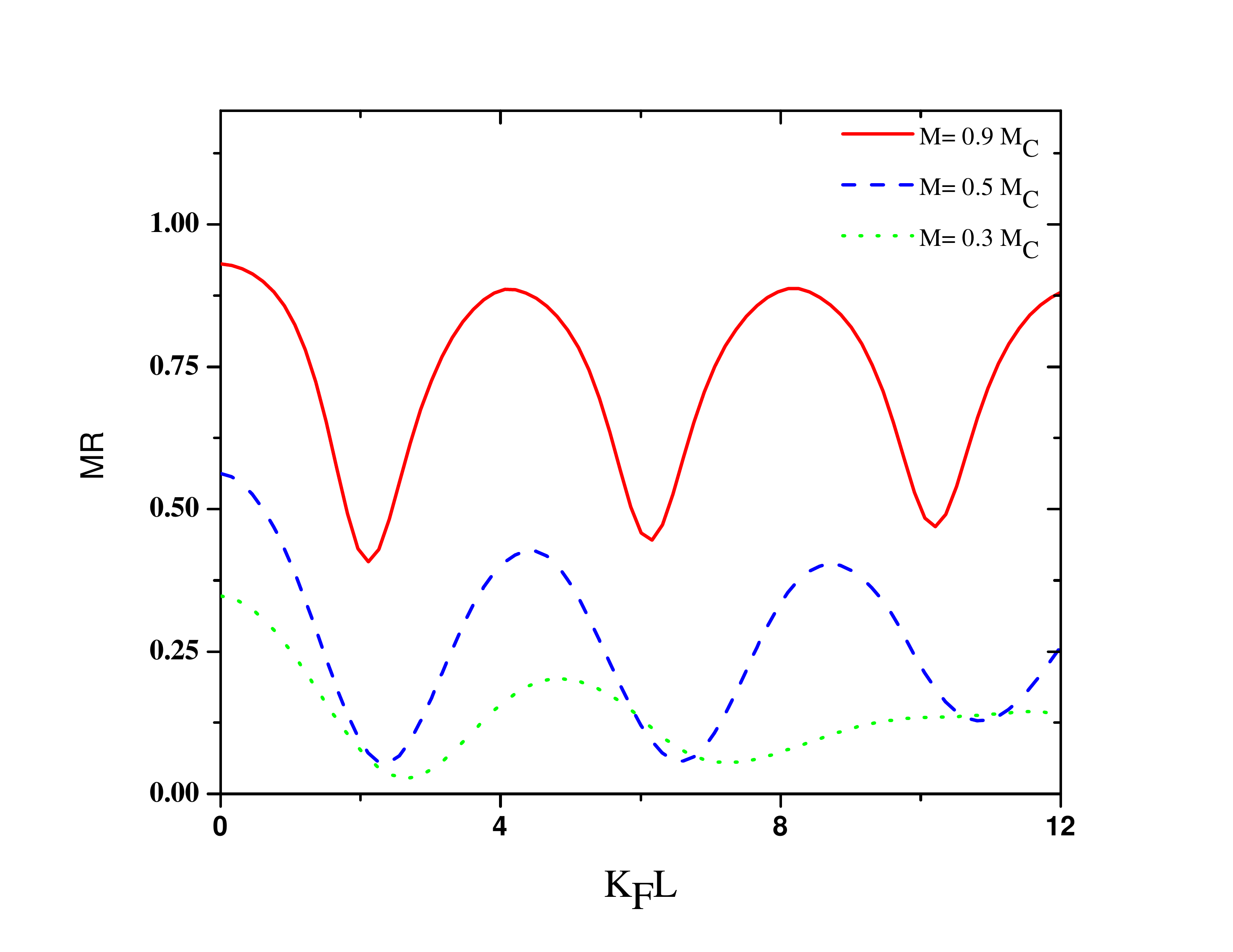}}
\caption{(Color online) The magnetoresistance of the F$\mid$N$\mid$F
junctions vs. width of the
 normal region ($K_FL=\frac{\mu}{\hbar v_F}L$) for three values of $M/M_C$. $eV$ the gate voltage of junction is
 fixed at $0.1\mu$.} \label{[I]Mr-kfl}
\end{figure}
\begin{figure}[tbp]
\centering{\includegraphics[width=0.8\columnwidth]{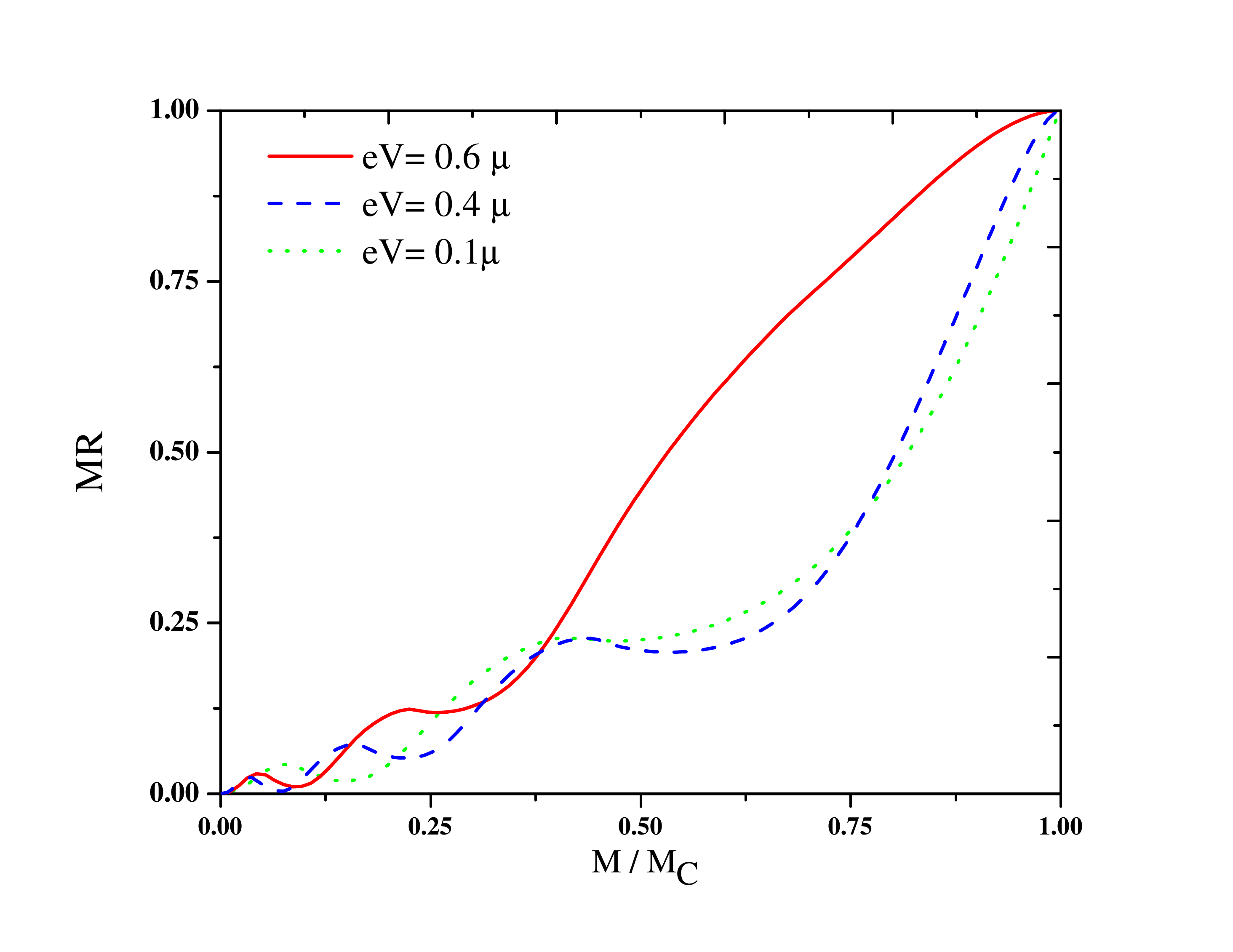}}
\caption{(Color online) The magnetoresistance of the F$\mid$N$\mid$F
junctions vs. $M/M_c$ for three values of gate voltage $eV$ and the
width of the junction is fixed at $K_FL$=10.} \label{[I]MR-Mc}
\end{figure}
This property is reason for some interesting physical properties
of the system\cite{Fu3}. For three-dimensional TI, strong (weak)
TI is defined for odd (even) number of chiral Dirac fermions on
surface\cite{Fu2}. In graphene, There are four Dirac cones in
momentum space \cite{Novoselov}, but the number of Dirac cones in
TI's edge are odd. So there are good unconventional properties
such as fractional quantum Hall effect, $\pi$ Berry's phases and
immunity to Anderson localization due to odd spin-Dirac on the
surface\cite{Hsieh,Xia,Qi,Schnyder}. On the other hand, studying
of transport properties is useful to understand the macroscopic
and applicatory properties. Some attractive physical properties
are seen when TI is on proximity of ferromagnet or
superconductor\cite{Essin,Franz,Akhmerov,linder1,linder2}.
 Also the tunneling magnetoresistance (TMR) effect has attracted much interest because
of crucial applications in spintronics \cite{Zutic}. In this paper
we investigate TMR and transport properties of the Dirac fermions on
the surface of topological insulator for F$\mid$N$\mid$F and
F$\mid$B$\mid$F junctions within the clean limit and low temperature
regime. We assume that the used topological insulator
 is made of $HgTe$ or $Bi_2Se_3$. The advantage of the mentioned
alloys is their single Dirac cone in Berillouin zone in which two
dimensional Dirac equation govern the Dirac fermions on surface of
the topological insulator\cite{Hsieh,Bernevig,Xia}. Due to the
exchange coupling, the ferromagnetism can be induced on the
surface of topological insulator when a ferromagnetic layer is
deposited on top of it\cite{Yokoyama}. We use two dimensional
Dirac Hamiltonian and derive wave functions in all regions. Using
boundary conditions we obtain analytical expressions for
reflection and transmission probabilities. Using the obtained
reflection and transmission probabilities, we calculate electronic
conductance of the two mentioned systems when the magnetization of
ferromagnetic layers are parallel and antiparallel. To investigate
properties of F$\mid$B$\mid$F junctions we use the thin barrier
approximation. The used barrier can be made by applying a
transverse voltage or doping\cite{Tanaka,Sengupta}. The main
concern is to investigate the magnetoresistance of the mentioned
systems and their differences.

\section{Tunneling Magnetoresistance in F$\mid$N$\mid$F Junctions}
\label{strain} On the surface of topological insulator Dirac
fermions are governed by two dimensional Dirac Hamiltonian that is
given by:
\begin{equation}
H = \hbar v_F \vec{\sigma}.\vec{k}-\mu I \label{dirac hamiltonian}
\end{equation}
where $\vec{\sigma}$ is vector of Pauli matrices in spin space and
I is identity matrix. Also, $v_F$ is Fermi velocity and $\mu$ is
chemical potential. When a ferromagnetic electrode with
magnetization {\bf m} is deposited on surface of the TI, due to
the proximity of TI and ferromagnetism, an additional term appear
into Eq.(\ref{dirac hamiltonian}) as follow\cite{Yokoyama}:
\begin{equation}
H_0 = \hbar v_F \vec{\sigma}.\vec{m}.\label{magnet
hamiltonian}\end{equation} We consider an F$\mid$N$\mid$F junction
in which thickness of normal region is $d$.
 We assume the TI is located in $xy$-plane and two ferromagnetic
electrodes is deposited on TI such that
$\mathbf{m}=\mathbf{\hat{y}}m_0(\theta(-x)\pm\theta(x-d))$ in
which +(--) refers to parallel (antiparallel) in magnetization
alignment of electrodes (See Fig. \ref{Model}). Here $\theta(x)$
is the well known step function. By solving the Eq. (\ref{dirac
hamiltonian}) for $x<0$ region, Dirac spinors read as:
\begin{equation}
\psi_{I}^{\pm}(x,y)=\frac{1}{\sqrt{2}}e^{(\pm ik_{x} x+ik_y
y)}\left(\begin{array}{c}
     1 \\
     \pm e^{\pm i \beta}\\
   \end{array}\right),
\label{Spinor I}
\end{equation}
where $k_x$ and $k_y$ are components of wave vector in $x$ and $y$
directions, respectively. Also the $\pm$ sings refer to the
directions of the $\hat{x}$-axis and $\beta$ is incidental angle
of the Dirac fermions which $\beta$ and $k_x$ are described as:
\begin{eqnarray}
\nonumber \beta&=&\arcsin{\frac{\hbar v_F (k_y+M)}{|\epsilon+\mu|}},\\
k_x&=&\sqrt{({\frac{\epsilon+\mu}{\hbar v_F}})^2-(k_y+M)^2},
\label{beta}
\end{eqnarray}
where $M=m_0/\hbar v_F$ depends on the exchange coupling and one
can tune the parameter via an external field\cite{Yokoyama}. For
normal region between $0<x<d$, Dirac spinors are given by:
\begin{equation}
\psi_{II}^{\pm}(x,y)=\frac{1}{\sqrt{2}}e^{(\pm ik_{x}^{\prime
\prime} x+ik_y y)}\left(\begin{array}{c}
     1 \\
     \pm e^{\pm i \alpha}\\
   \end{array}\right)
\label{Spinor II}
\end{equation}
in which $k_x^{\prime \prime}$ is $x$-component of the wave vector
and $\alpha$ is angle of incidental Dirac fermions and are defined
as:
\begin{eqnarray}
\nonumber \alpha&=&\arcsin{\frac{\hbar v_F k_y}{|\epsilon+\mu|}}\\
k_x^{\prime \prime}&=&\sqrt{({\frac{\epsilon+\mu}{\hbar
v_F}})^2-k_y^2} \label{alpha}
\end{eqnarray}

In region $ x>d $, there are two choices, parallel or antiparallel
directions for magnetization of the electrodes. When we make our
system with parallel magnetization, all spinors are similar to Eq.
(\ref{Spinor I}) but for antiparallel case the spinors read:
\begin{equation}
\psi_{III}^{\pm}(x,y)=\frac{1}{\sqrt{2}}e^{(\pm ik_{x}^{\prime}
x+ik_y y)}\left(\begin{array}{c}
     1 \\
     \pm e^{\pm i \beta^{\prime}}\\
   \end{array}\right)
\label{Spinor III}
\end{equation}
in which $\beta^{\prime}$ and $k_x^{\prime}$ are defined as:
\begin{eqnarray}
\nonumber \beta^{\prime}&=&\arcsin{\frac{\hbar v_F (k_y-M)}{|\epsilon+\mu|}}\\
k_x^{\prime}&=&\sqrt{({\frac{\epsilon+\mu}{\hbar v_F}})^2-(k_y-M)^2}
\label{beta_prime}
\end{eqnarray}
As seen in Eqs.(\ref{beta}, \ref{beta_prime}), for antiparallel
magnetization case when $M$ is larger than its critical value
namely $M_C=|\epsilon+\mu|/\hbar v_F$, the wave function in Eq.
(\ref{Spinor III}) decays. For an incidental particle from $x<0$
into  the F$\mid$N interface at $x=0$, the transverse wave vector
$k_y$ and excitation energy $\epsilon$ are conserved in scattering
process. The Dirac spinors within each region must satisfy the
following boundary conditions at $x=0$:
\begin{equation}
\psi_{I}^{+}+r {\psi}_{I}^{-}=a_1{\psi}_{II}^{+}+a_2
{\psi}_{II}^{-}. \label{B.C I}
\end{equation}
and at $x=d$ :
\begin{equation}
t \psi_{I}^{+}=a_1{\psi}_{II}^{+}+a_2 {\psi}_{II}^{-}. \label{B.C
II}
\end{equation}

By solving Eqs.\ref{B.C I} and \ref{B.C II}, transmission
probability $T=|t|^2$ for parallel and antiparallel magnetization
directions are derived as:
\begin{equation}
T_{AP}=\frac{\cos^2\beta \cos^2 \alpha
}{A\cos^2(\frac{\beta+\beta^\prime}{2}) \cos^2 \alpha+
B\ \ [\cos(\frac{\beta-\beta^\prime}{2})-\sin\alpha
\sin(\frac{\beta+\beta^\prime}{2})]^2}\label{T_P}
\end{equation}
\begin{equation}
T_{P}=\frac{\cos^2\beta \cos^2 \alpha
}{A \cos^2(\beta) \cos^2 \alpha+
B\ \ [1-\sin\alpha \sin(\beta)]^2}
\label{T-AP}
\end{equation}
where $A=\cos^2(\frac{1}{\sqrt{2}}k_x d)$ and
$B=\sin^2(\frac{1}{\sqrt{2}}k_x d)$. The P and AP indexes stand
for parallel and antiparallel magnetizations. The Eqs.\ref{T_P}
and \ref{T-AP} are the main results of this section. Using the
transmission probabilities and conductance formula\cite{Sengupta},
one can calculate the conductances of the mentioned structures
containing parallel and antiparallel magnetizations,
\begin{equation}\label{conductance}
G(\epsilon)=\frac{G_{0}}{2} \int_{-\pi/2}^{\pi/2} T\cos\beta{\rm
d}\beta
\end{equation}
here $G_0=N(\epsilon)W e^2 / \pi \hbar^2 v_F$ is normal conductance
and $N(\epsilon)=|\mu+eV|/2 \pi(\hbar v_F)^2$ is
density of state of Dirac fermions and $W$ is width of the junction.\\
The conductances of the parallel and antiparallel magnetization
are plotted versus $k_F L$ in Fig.\ref{[I]GP-GAP-kfL}, here
$k_F=\mu/\hbar v_F$ is Fermi vector of the system. The oscillatory
behavior of both conductances are results of coherent interference
of the Dirac fermions at two interfaces. By using the conductances
of parallel and antiparallel magnetization, one can calculate the
magnetoresistance of the system which is defined as $MR =100\%
\times (G_P -G_{AP})/G_P$. This quantity may have many interesting
and important applications in spintronic industry, in similarity
with the Ref. \cite{Zutic}. The plotted magnetoresistance as a
function of $k_FL$ is shown in Fig.\ref{[I]Mr-kfl} which shows
oscillatory behavior for all of the gate voltages as we expected.
This oscillatory behavior is a result of oscillatory behavior in
conductances  for parallel and antiparallel magnetizations. The
plotted magnetoresistance as a function of $M$ can be seen in
Fig.\ref{[I]MR-Mc} that is also one of the main results of this
section. By increasing $M$, the magnetoresistance enhance to reach
the perfect value in $M_C$. Although for small values of $M$ the
magnetoresistance oscillates, the quantity for large values of
$M$, shows a smooth behavior. Now we proceed to present an
investigation of the FI/BI/FI junction.
\begin{figure}[tbp]
\centering{\includegraphics[width=0.8\columnwidth]{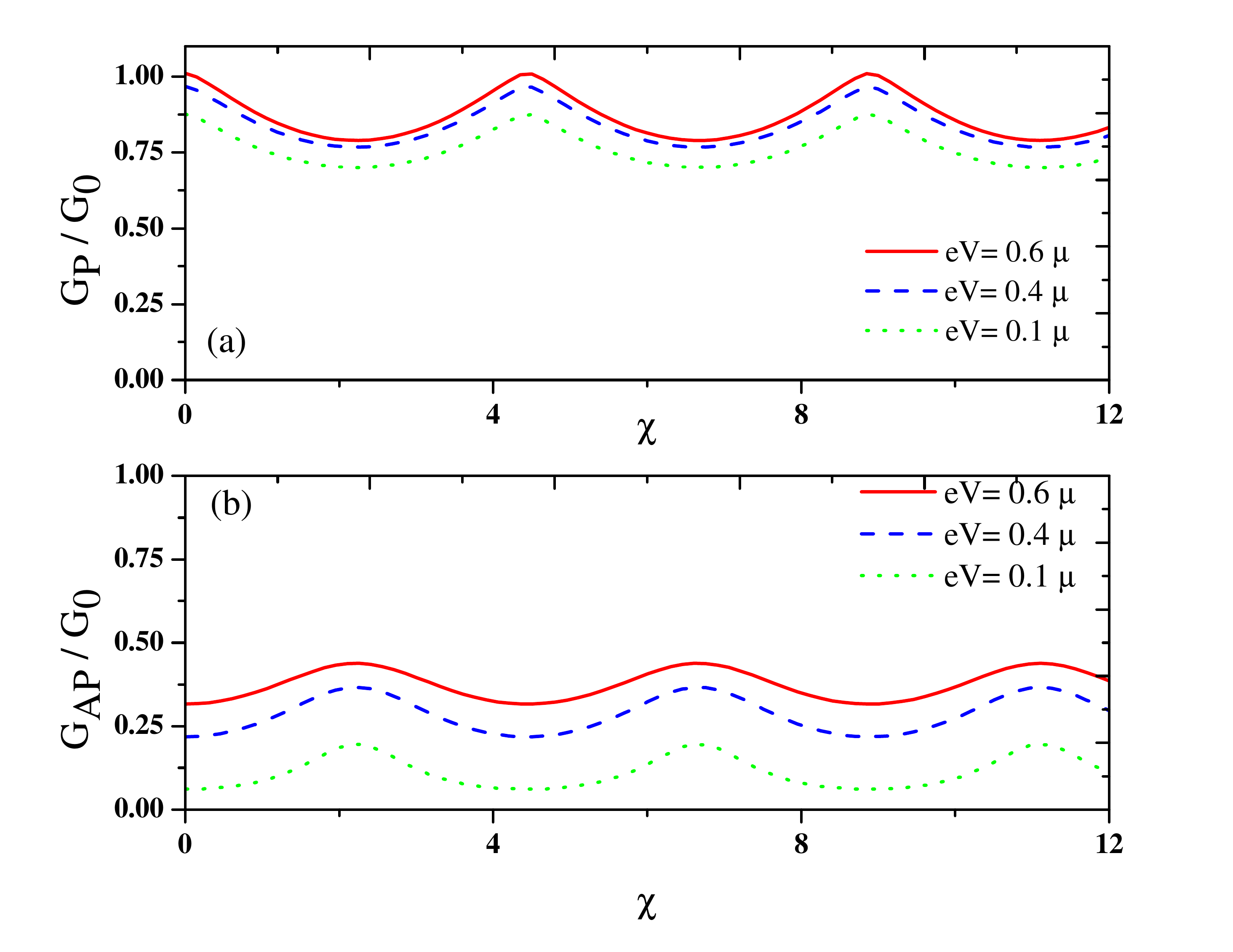}}
\caption{(Color online) The electronic conductances of
F$\mid$B$\mid$F junctions within the thin barrier approximation vs.
strength of barrier $\chi=V_0 d /\hbar v_F$ for three values of gate
voltage $eV$. Here $M/M_C$ is fixed at 0.9.} \label{[II]Gp-GAP-xi}
\end{figure}
\begin{figure}[tbp]
\centering{\includegraphics[width=0.8\columnwidth]{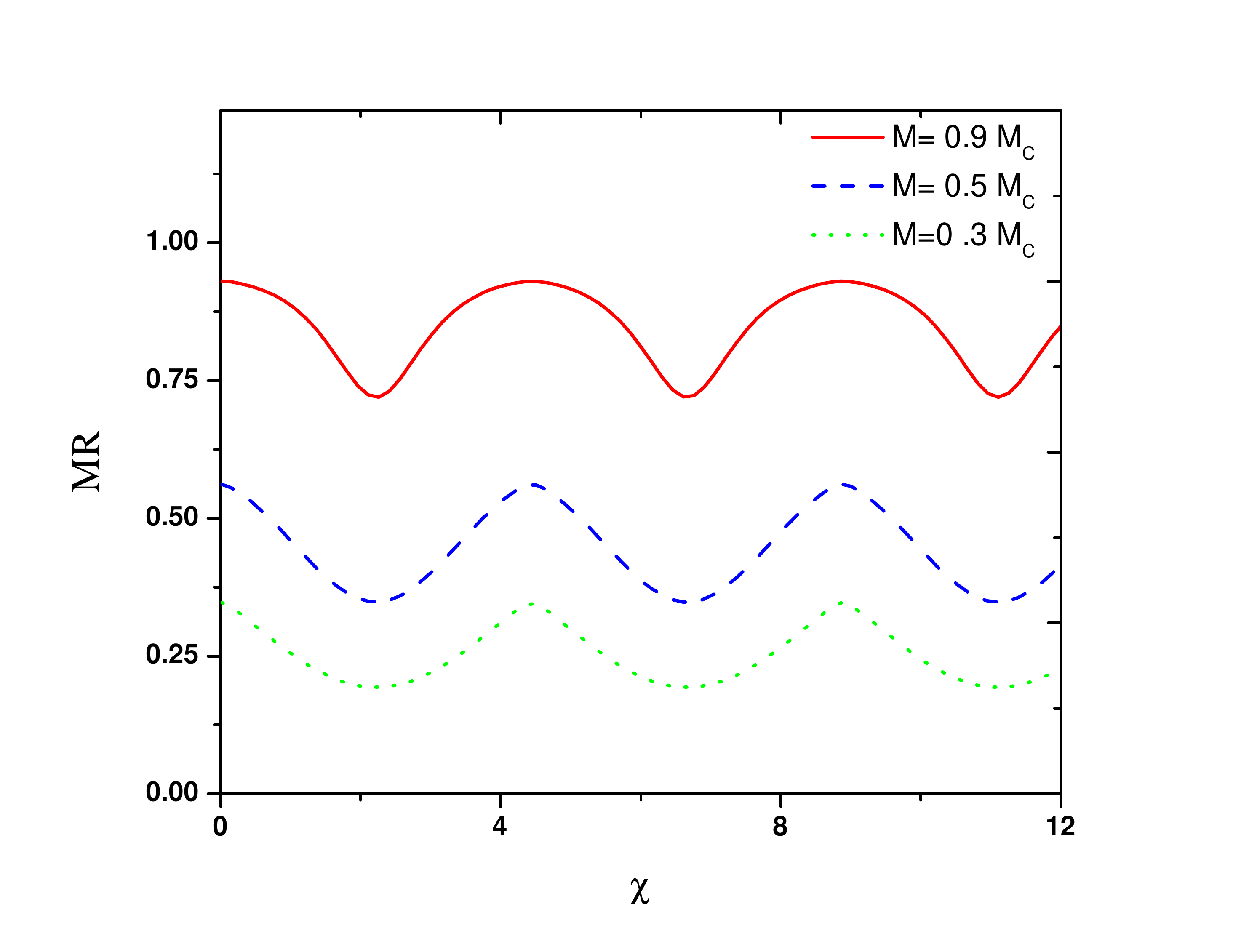}}
\caption{(Color online) The magnetoresistance of the F$\mid$B$\mid$F
junctions within the thin approximation vs. strength of barrier
$\chi=V_0 d /\hbar v_F$ for three values of $M/M_C$ and fixed gate
voltage $eV=0.1\mu$.} \label{[II]MR-xi}
\end{figure}
\begin{figure}[tbp]
\centering{\includegraphics[width=0.8\columnwidth]{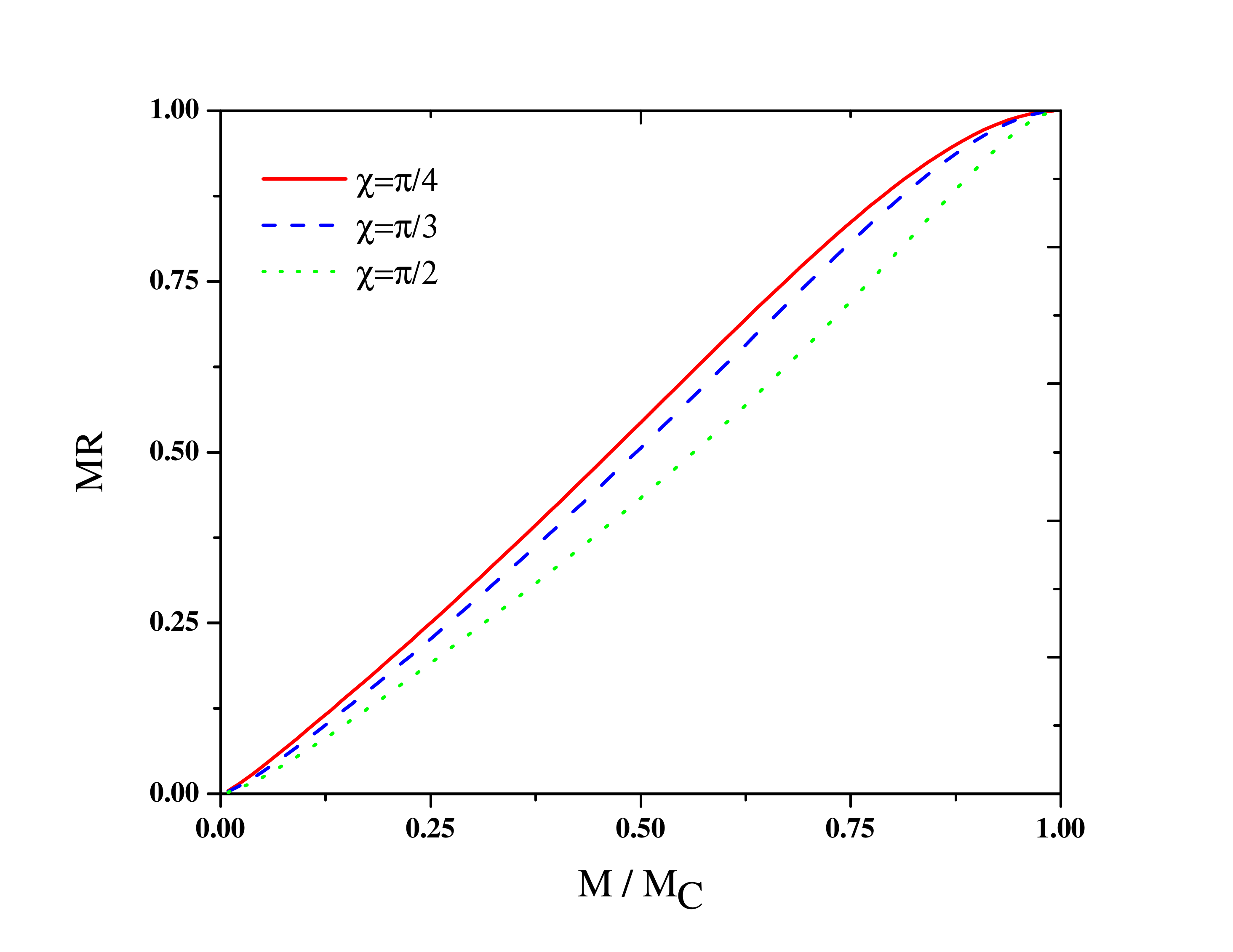}}
\caption{(Color online) The magnetoresistance of the F$\mid$B$\mid$F
junctions within the thin approximation vs. $M/M_C$ for three values
of strength of barrier $\chi=V_0 d /\hbar v_F$ and fixed gate
voltage $eV=0.1\mu$.} \label{[[II]MR-Mc}
\end{figure}
\section{Tunneling Magnetoresistance in F$\mid$B$\mid$F Junctions}
\label{theory} Now we focus our attention to F$\mid$B$\mid$F
junctions. For producing such structures one can apply a
transverse gate voltage into the normal region and make it as a
barrier for moving Dirac fermions from left ferromagnet electrode
to interface (See Fig.\ref{Model}, part (b)). By applying the
barrier voltage, another additional term added to the Dirac
Hamiltonian Eq. (\ref{dirac hamiltonian}) for normal region and
change it as:
\begin{equation}
\label{barrier hamiltonian} H = \hbar v_F \vec{\sigma}.\vec{k}+V_0 I
-\mu I.
\end{equation}
Dirac fermion states within the barrier region are:
\begin{equation}
\psi_{B}^{\pm}(x,y)=\frac{1}{\sqrt{2}}e^{(\pm ik_{B,x}^{\prime
\prime} x+ik_y y)}\left(\begin{array}{c}
     1 \\
     \pm e^{\pm i \theta}\\
   \end{array}\right)
\label{Spinor B}
\end{equation}
where $\theta$ and $k_{B,x}^{\prime \prime}$ are defined as:
\begin{eqnarray}
\nonumber \theta&=&\arcsin{\frac{\hbar v_F k_y}{|\epsilon-eV_0+\mu|}}\\
k_{B,x}^{\prime \prime}&=&\sqrt{({\frac{\epsilon-eV_0+\mu}{\hbar
v_F}})^2-k_y^2} \label{theta}
\end{eqnarray}
We are interested in the influence of thin barrier regime on the
magnetoresistance. In this limit, width of the junction $d$
approaches to zero and barrier voltage $V_0$ move to large values
simultaneously so that the barrier strength $\chi = eV_0d/\hbar
v_F $ remains constant. In this limit $\theta\approx 0$ and
$k_{B,x}^{\prime \prime}d\approx\chi$.
 Substituting the present Dirac spinors Eq. (\ref{Spinor B}) into the
boundary conditions Eqs.\ref{B.C I} and \ref{B.C II} , we have derived
the transmission probabilities for parallel (P label) and
antiparallel (AP label) structures within the thin barrier regime as
below:
\begin{eqnarray}
&T_{AP}&=\frac{\cos^2\beta}{A'\cos^2(\frac{\beta+\beta^\prime}{2})+
B' \ \ [\cos(\frac{\beta-\beta^\prime}{2})-
\sin(\frac{\beta+\beta^\prime}{2})]^2 }\label{[II]T-AP},\\
&T_{P}&=\frac{\cos^2\beta}{A'\cos^2(\beta)+ B'\ \
[1-\sin(\beta)]^2}. \label{[II]T-P}
\end{eqnarray}
where $A'=\cos^2(\frac{\chi}{\sqrt{2}})$ and
$B'=\sin^2(\frac{\chi}{\sqrt{2}} )$. The above obtained analytical
expressions for the transition probabilities of parallel and
antiparallel structures are other results of the present paper. As
seen in Eq. (\ref{[II]T-P}),
 for an incidental particle with normal direction to the
interface, the transmission probability is equal to unity. This
situation is similar to the Klein tunneling in quantum
relativistic theory \cite{Klein}. To calculate the conductance of
the mentioned parallel and antiparallel structure within the thin
barrier regime, one should use the obtained transmission
probabilities (Eqs.\ref{[II]T-AP},\ref{[II]T-P}) in Eq.
(\ref{conductance}).
 The plotted conductances for both parallel and
antiparallel magnetization are shown in Fig.\ref{[II]Gp-GAP-xi} as a
function of the barrier strength $\chi$.
 Changing the gate voltage doesn't make any changes in the positions of the
extremum in conductances of parallel and antiparallel magnetization.
 The maximum of parallel conductance has $\pi/2$ phase difference with respect
to antiparallel case. Unlike damping oscillatory behavior of MR
versus length of the normal region $K_F L$ in F$\mid$N$\mid$F case
(See Fig.\ref{[I]Mr-kfl}), here the oscillatory behavior of the
conductance leads to harmonic oscillations in MR, as shown in
Fig.\ref{[II]MR-xi}. We plotted the magnetoresistance for this set
up as a function of $M$, as shown in Fig.\ref{[[II]MR-Mc}, which
approximately shows linear behavior. This linear manner in
magnetoresistance for F$\mid$B$\mid$F junction seems to be more
applicatory than the discussed behavior of magnetoresistance in
F$\mid$N$\mid$F junctions.
 By tuning the gate voltage and
exchange coupling due to the ferromagnetic electrodes, we can access
to an arbitrary value in magnetoresistance which is useful in spintronics.
\section{Conclusions}
\label{conclude} In summary, we have investigated the electronic
transport properties of topological insulator-based
ferromagnetic/normal/ferromagnetic (F$\mid$N$\mid$F) and
topological insulator-based ferromagnetic/barrier/ferromagnetic
F$\mid$B$\mid$F junctions on the surface of topological insulator
in low temperature and clean regimes. We assumed the two
magnetizations are in-plane and deposited on the surface of
topological insulator. We have derived analytical expressions for
transmission probabilities of two different magnetization
directions, parallel and antiparallel for F$\mid$N$\mid$F and
F$\mid$B$\mid$F junctions. To investigate electronic properties of
the F$\mid$B$\mid$F junction, in our calculations we assumed that
$d\approx 0$ and $V_0\rightarrow \infty$ simultaneously. This
approximation is well known as thin barrier approximation. Using
the derived transition probabilities, conductances and
magnetoresistances of the systems were calculated. We have found
that the very important quantity of such junctions
(magnetoresistance) shows different behavior by increasing the
magnetization strength of ferromagnetic region for F$\mid$N$\mid$F
and F$\mid$B$\mid$F junctions. Although magnetoresistance of the
two junctions can be tuned from values near $0$ up to $100\%$,
magnetoresistance for F$\mid$N$\mid$F junctions shows oscillatory
behavior vs. $M/M_C$ the magnetization strength, variation of the
quantity vs. $M/M_C$ for F$\mid$B$\mid$F junctions behave like a
very smooth enhancement.
\section{Acknowledgments}
The authors appreciate very useful and fruitful discussions with
Jacob Linder. Rasoul Ghasemi also is appreciated for helpful points
in revision. We would like to thank the Office of Graduate Studies
of Isfahan University.

\end{document}